\documentclass[12pt]{article}

\usepackage{graphicx,amssymb,url,mathrsfs, amsmath}
\usepackage{eucal}
\usepackage{wrapfig}
\usepackage{boxedminipage}
\usepackage{setspace}
\usepackage{subfigure}
\usepackage[dvipdfm]{hyperref}

\def\l{\left}
\def\r{\right}

\begin{document}

\begin{titlepage}

\begin{flushright}
{\tt hep-th/...}
\tt {FileName:....tex} \\
{\tt \today}

\end{flushright}
\vspace{0.5in}

\begin{center}
{\large \bf Jet quenching in shock waves}\\
\vspace{10mm}
Michael Spillane, Alexander Stoffers and Ismail Zahed\\
\vspace{5mm}
{\it \small Department of Physics and Astronomy, Stony Brook University, Stony Brook, NY 11794, USA}\\
          \vspace{10mm}
{\tt \today \vspace{1cm}}
\end{center}
\begin{abstract}
We study the propagation of an ultrarelativistic light quark jet inside a shock wave using the holographic principle. The maximum stopping distance and its dependency on the energy of the jet is obtained. 
\end{abstract}
\end{titlepage}

\renewcommand{\thefootnote}{\arabic{footnote}}
\setcounter{footnote}{0}

%\tableofcontents
%\newpage

%%%%%%%%%%%%%%%%%%%%%%%%%%%%%%%%%%%%%%%%%%%%%%%%%%%%%%%%%%%%%%%%%%%%%%%%%%%
%%%%%%%%%%%%%%%%%%%%%%%%%%%%%%%%%%%%%%%%%%%%%%%%%%%%%%%%%%%%%%%%%%%%%%%%%%%
%%%%%%%%%%%%%%%                 BODY                    %%%%%%%%%%%%%%%%%%%
%%%%%%%%%%%%%%%%%%%%%%%%%%%%%%%%%%%%%%%%%%%%%%%%%%%%%%%%%%%%%%%%%%%%%%%%%%%
%%%%%%%%%%%%%%%%%%%%%%%%%%%%%%%%%%%%%%%%%%%%%%%%%%%%%%%%%%%%%%%%%%%%%%%%%%%

\section{Introduction}

A shock wave can be interpreted as a dense, non-equilibrated medium and the collision of two shock waves has been used to model heavy ion collisions and to study the formation of a thermalized, strongly coupled medium, see for example \cite{Gubser:2008pc, Lin:2009pn,Kovchegov:2009du,Chesler:2010bi}. To gain insight into the energy loss at the early stages of the collision, we use the holographic principle to study the propagation of an ultrarelativistic light quark jet in a shock wave. 
The propagation of probe quarks in a strongly coupled medium can be accessed by studying strings in a suitable AdS background and the energy loss of a heavy quark inside a shock wave is discussed in \cite{Horowitz:2009pw}.
While the bulk metric describing the late time asymptotics of a relativistic, hydrodynamic medium is known  \cite{Janik:2005zt}, an analytic solution for the metric valid at times shortly after the collision of two shock waves is still missing; see however \cite{Grumiller:2008va}.
A setup dual to an ultrarelativistic light quark jet moving though an equilibrated medium was proposed in \cite{Chesler:2008wd, Chesler:2008uy}. In particular, the energy scaling of the maximum stopping distance was shown to be $\l(\Delta x (E)\r)_{max} \propto \frac{1}{T}\l(\frac{E}{T\sqrt{\lambda}}\r)^{1/3}$.
We will follow the reasoning of this approach to evaluate the energy dependence of the maximum stopping distance for an ultrarelativistic light quark jet in a shock wave and neglect any gravitational back reaction \cite{Shuryak:2011ge}. For another approach to jet quenching in a holographic frame work, see \cite{Arnold:2010ir, Arnold:2011qi}.

In order to evaluate the maximum stopping distance for two light quark jets flying back to back, Chesler et. al. \cite{Chesler:2008wd, Chesler:2008uy} consider an open string in an asymptotically $AdS_5$ background whose bulk is filled by a D7 flavor brane. An early approach to jet quenching by using light rays in bulk can be found in \cite{Sin:2004yx}; see also \cite{Gubser:2008as, Hatta:2007cs,Hatta:2008tx,Stoffers:2011fx}. At late times after the initialization of the jet, the string can be approximated by a trailing null string. The trajectory of endpoints of the string is well approximated by a null geodesic and the jets are interpreted to diffuse into the medium once the string endpoints reach the horizon of the black hole.

\section{Jet quenching using light rays in bulk}

Motivated by the observation, that the shape of a trailing and falling string can be captured by studying light rays propagating in bulk, \cite{Hatta:2007cs, Hatta:2008tx, Stoffers:2011fx}, we consider the propagation of a nearly massless particle in bulk to get an estimate for the maximum stopping distance for a quark jet within a shock wave. 
We consider a particle following a null geodesic in the background dual to a shock wave solution. Since the metric in (\ref{metric}) yields a conserved momentum and energy, we fix the resulting constants of motion
\begin{eqnarray}
\dot{x}=\frac{dx}{dt}=\frac{\Pi_x}{\Pi_t}f=v f \ 
\end{eqnarray}
to describe a quark at the boundary with velocity $v$. The null geodesic reads  
\begin{eqnarray}
\dot{z}=\frac{dz}{dt} = \sqrt{f-\dot{x}^2} \ ,
\end{eqnarray} 
and for $\dot{z}$ to be real, the bulk coordinate is confined to $z^4<1/\mu$. The distance $\Delta x$ the quark travels is given by
\begin{eqnarray}
\Delta x = \int_0^{\mu^{-1/4}} dz \frac{\dot{x}}{\dot{z}} = \int_0^{\mu^{-1/4}} dz \frac{v \sqrt{f}}{\sqrt{1-v^2f}} = \int_0^{\mu^{-1/4}} dz \frac{v\sqrt{f}}{\sqrt{\frac{1}{\gamma^2}+v^2 \mu z^4}}
\end{eqnarray}
For $v\simeq 1$ and finite $\frac{1}{\gamma}$, the energy dependence of the maximum penetration depth reads
\begin{eqnarray} \label{penetrationdepth}
\l(\Delta x\r)_{max} \approx \sqrt{\gamma} \mu^{-1/4} \approx \mu^{-1/4} \sqrt{E/M(\mu)} \ ,
\end{eqnarray} 
where $E=\gamma M(\mu)$ is the energy of  an ultrarelativistic quark with an in-medium mass $M(\mu)$ zipping through a plasma. Following \cite{Horowitz:2009pw, Albacete:2008ze} we can relate the coefficient $\mu$ to the typical particle momentum $\Lambda$ of the shock wave medium boosted along $x^3$ with Lorentz factor $\gamma_{\parallel}$ by
\begin{eqnarray}
\mu = \pi^2 \Lambda^4 \gamma_{\parallel}^2 \ .
\end{eqnarray}
Since the maximum stopping distance (\ref{penetrationdepth}) is governd by the UV behavior of the null geodesic and the geometries for a shock wave, (\ref{metric}), and thermal AdS$_5$ coincide near the boundary, the energy dependence of the stopping distance in a shock wave and in thermal AdS are the same, compare \cite{Stoffers:2011fx}.

\section{Stopping distance for a ultrarelativistic light quark jet in a shock wave}

The metric dual to a shock wave with energy momentum tensor $< T_{--}> = \frac{N_c^2}{\pi^2} \mu \theta (x^-)$ reads \cite{Horowitz:2009pw, Janik:2005zt}
\begin{eqnarray} \label{metric}
ds^2=G_{MN}dx^M dx^N=\frac{R^2}{z^2} \l(-f(z) dt^2 + dx^2 + dz^2 \r) \ ,
\end{eqnarray}
with the AdS radius $R$, $x_\perp(t,z)\equiv x(t,z)$, $f(z)=1-\mu \theta(x^-) z^4$ and $x^-=\frac{1}{\sqrt{2}} (t-x^3)$. Since we are interested in ultrarelativistic dynamics in the $x_\perp$-direction, the longitudinal components proportional to $x^3$ in (\ref{metric}) are suppressed and $\theta (x^-)=1$. A light ray can escape from the region $z \le \mu^{-\frac{1}{4}}$ and the surface at $z=\mu^{-\frac{1}{4}}$ is not a true horizon \cite{Horowitz:2009pw}. 
Null geodesics in the ($t,z$)-plane follow
\begin{eqnarray}
\dot{x}_{geo}^2&\equiv & \l(\frac{dx}{dt}\r)_{geo}^2 = f^2\\ \label{nullgeodesicforz}
\dot{z}_{geo}^2 &\equiv & \l(\frac{dz}{dt}\r)_{geo}^2 = f(1-f) 
\end{eqnarray}
and the trajectory in bulk is given by
\begin{eqnarray} \label{nullgeodesic}
x_{geo}'^2 \equiv  \l(\frac{dx}{dz}\r)_{geo}^2 = \frac{f}{1-f} \ .
\end{eqnarray}
The length of the path of the null geodesic in $x-$direction increases the closer the initial radial position is to the boundary; close to the boundary at some $z=z_*\simeq0$, (\ref{nullgeodesic}) implies
\begin{eqnarray}
x(z_*)\simeq \frac{1}{ \sqrt{\mu} z_*} \ .
\end{eqnarray}
As shown in \cite{Chesler:2008wd, Chesler:2008uy}, the string is at late times but long before the jet diffuses into the medium well approximated by a trailing null string solution with the endpoints following null geodesics. We adopt this assumption and approximate the string by
\begin{eqnarray} \label{Ansatz}
x(t,z)&=&x_0(t,z)+ \delta x(t,z) + \mathcal{O}\l( (\delta x)^2\r) \\
\mathcal{Z}(t) &=& \mathcal{Z}_0(t) + \delta \mathcal{Z}(t) + \mathcal{O}\l( (\delta \mathcal{Z})^2\r)\ ,
\end{eqnarray}
with $x_0(t,z)$ a trailing null string solution and the trajectory of the endpoints, $\mathcal{Z}_0(t)$, following null geodesics as we will now show. In the static gauge the Nambu-Goto action in the background (\ref{metric})
\begin{eqnarray}
S = \frac{\sqrt{\lambda}}{2\pi R^2} \int dt \ dz \sqrt{-g} = \frac{\sqrt{\lambda}}{2\pi} \int dt \ dz  \frac{1}{z^2} \sqrt{f+ f x'^2 - \dot{x}^2} \ ,
\end{eqnarray}
with the t'Hooft coupling $\lambda$ results in the following solution for a string of the form $x_0(t,z)=vt + \xi(z)$
\begin{eqnarray} \label{eom}
\xi'^2= \frac{f-v^2}{f} \frac{1}{\l(\frac{R^4}{z^4}\r) \frac{f}{c}-1} \ .
\end{eqnarray}
The integration constant $c$ is inversly proportional to the on-shell action density $\sqrt{-g}$. 
The endpoint of the open string follows a curve $\mathcal{Z}_0(t)$ fixed by the boundary conditions
\begin{eqnarray} 
G_{AB} \frac{d X^A}{dt}\frac{d X^B}{dt}=0 \\
G_{AB} \frac{d X^A}{dt}\frac{\partial X^B}{\partial z}=0 \ ,
\end{eqnarray}
($X^A=(t,x,z)$) which are equivalent to
\begin{eqnarray} \label{boundarycondition1}
\xi'^2 &=& \frac{v^2-f}{f}  \\ \label{boundarycondition2}
\dot{\mathcal{Z}}_0^2 &=& \frac{f(v^2-f)}{v^2} \ .
\end{eqnarray}
To match the boundary condition (\ref{boundarycondition1}) on to the string, (\ref{eom}), $\frac{1}{c}=0$, which means the string is a null string with vanishing action density. In order for the string to trail behind the endpoint, the sign of the square root in (\ref{boundarycondition1}) is chosen such that $\xi'=-\sqrt{\frac{v^2-f}{f}}$. For a light quark with $v\simeq 1$, the endpoint of the string, (\ref{boundarycondition2}), is well approximated by a null geodesic, compare (\ref{nullgeodesicforz}). \\
The energy of the string at a time $t_*$ is given by \cite{Chesler:2008wd,Chesler:2008uy}
\begin{eqnarray} \label{energy1}
E_*=\frac{\sqrt{\lambda}}{2 \pi R^2} \int_{z_*}^{\infty} dz \frac{G_{tB}}{\sqrt{-g}} \l( \l( \dot{X} \cdot X'\r) X'^{B}- \l(X' \cdot X'\r) \dot{X}^B \r)  \ ,
\end{eqnarray}
where we have assumed that the flavor brane fills the whole AdS space, effectively setting the rest mass of the light quark to zero. The maximum stopping distance is related to the UV behavior of the trailing string and the trajectory of the null geodesic. Thus, introducing a finite mass by extending the flavor brane from $z=0$ to some finite IR cutoff $z=z_m$ will not affect the maximum stopping distance.
Since the energy of the null string $x_0(t,z)$ is not finite $(\sqrt{-g}=0)$, we need to evaluate the perturbation $\delta x$, (\ref{Ansatz}), to obtain a finite energy configuration. The linearized equations of motion are
\begin{eqnarray} \label{lineareom}
\delta \ddot{x} + f \mu z^4  \delta x'' + 2\sqrt{f \mu} \ z^2 \delta \dot{x}' + \frac{4 \mu^{\frac{3}{2}}z^5}{\sqrt{f}} \delta \dot{x}+ 2\frac{z^3}{\mu} \delta x' =0 \ ,
\end{eqnarray}
and the unique solution reads
\begin{eqnarray} \label{deltax}
\delta x(t,z) = \phi(\alpha(t,z)) - \l(\frac{\mu^{-\frac{1}{4}}}{z} + \mathcal{O}(z^3) \r) \psi(\alpha(t,z)) \ ,
\end{eqnarray}
with $\alpha(t,z)=\int^z \frac{d \tilde{z}}{\tilde{z}^2 \sqrt{1-\mu \tilde{z}^4}}-t$ and arbitrary functions $\phi(\alpha)$, $\psi(\alpha)$.
To first order in $\delta x$, the energy of one jet is given by 
\begin{eqnarray} \label{energy2}
E_* = \frac{\sqrt{\lambda}}{2 \pi} \int_{z_*}^{\infty} dz \frac{f}{z^2} \frac{1}{\sqrt{2 \psi(\alpha(t_*,z))}} \ .
\end{eqnarray}
In order to minimize the energy, $\psi$ has to be maximized. For high energetic jets with $z_*$ close to the boundary, $x$ scales as $x(z\simeq0) \propto z^3$, (\ref{boundarycondition1}), and for $\delta x$ to be a small perturbation, the maximum scaling of $\psi$ is $\psi(z\simeq 0) = \mu z^4 \tilde{\psi}(z) + \mathcal{O}(z^5)$, compare (\ref{deltax}), with $\tilde{\psi}$ finite at the boundary. Integrating the trajectory of the endpoints, (\ref{nullgeodesic}), from $z_*$ up to the horizon, yields the maximum stopping distance to leading order in $\mu^{1/4}z_*$ 
\begin{eqnarray}
\l(\Delta x\r)_{max}=\frac{1}{\sqrt{\mu}\ z_*} \ .
\end{eqnarray}
Using (\ref{energy2}) we can relate the position of the endpoint, $z_*$, at a given time $t_*$ to the energy of the jet and obtain
\begin{eqnarray}
\l(\Delta x (E_*)\r)_{max}&=&\frac{1}{\sqrt{\mu}} \l(\frac{2\pi \sqrt{\mu} E_*}{\sqrt{\lambda} \mathcal{J} }\r)^{\frac{1}{3}} \\
&=&\frac{1}{\gamma_{\parallel}^{\frac{2}{3}}}\l(\frac{2 E_*}{\sqrt{\lambda} \mathcal{J} \pi \Lambda^4}\r)^{\frac{1}{3}}
\end{eqnarray}
with $\mathcal{J}= \int_1^{\infty} dw \frac{1}{\sqrt{2 \tilde{\psi}(w)}}$. 
The maximum stopping distance scales as $E_*^{1/3}$, which we recognize as the same dependency as for the thermalized, static medium, \cite{Chesler:2008uy}. The stopping distance is sensitive to the relative velocity of the shock wave in the lab frame. 

\section{Conclusion}

The energy scaling of the maximum stopping distance for an ultrarelativistic light quark jet in a boosted shock wave agrees with the result obtained in a static, thermalized medium. This is not a surprise: In order to obtain $\l(\Delta x (E_*)\r)_{max}$, we need 1) the trajectory of the null geodesic in the (x-z) plane close to the boundary and 2) the scaling of the trailing string solution close to its endpoint. Both are dominated by the geometry of the space near the boundary: 1) The null geodesic only depends on the geometry of the space. 2) We can find the trailing string solution by looking at light rays in bulk \cite{Hatta:2008tx, Stoffers:2011fx}. For an ultrarelativistic quark, the string endpoint is close to the boundary, $z_* \simeq 0$, and the maximum scaling of the string solution dictated by the geometry of the space near the boundary. Since the AdS-Schwarzschild metric describing a static, thermal medium, $ds^2=\frac{R^2}{z^2} [-(1-\frac{z^4}{z_H^4}) dt^2 + dx^2 + \frac{1}{(1-\frac{z^4}{z_H^4})}dz^2 ]$, has the same structure near $z=0$ as the shock wave metric, (\ref{metric}), the energy dependence of the stopping distance agrees in both backgrounds.

\end{document}